\documentclass[%
reprint,
 amsmath,amssymb,
 aps, physrev,
]{revtex4-2}

\usepackage{graphicx}
\usepackage{dcolumn}
\usepackage{bm}
\usepackage{comment}
\usepackage[english]{babel}

\usepackage[utf8]{inputenc}
\usepackage[T1]{fontenc}

\usepackage{helvet}

\usepackage{caption}
\usepackage{capt-of}  
\usepackage[colorlinks=true,
            linkcolor=blue,    
            citecolor=blue,      
            urlcolor=blue         
           ]{hyperref}
\usepackage[dvipsnames]{xcolor}  

\begin{document}

\title{Dynamic Modulation of Long-Range Photon–Magnon Coupling}

\author{Alban Joseph}
\email{a.joseph.2@research.gla.ac.uk}
\affiliation{James Watt School of Engineering, Electronics \& Nanoscale Engineering Division, University of Glasgow, Glasgow, G12 8QQ, United Kingdom}

\author{Mawgan A. Smith}
\affiliation{James Watt School of Engineering, Electronics \& Nanoscale Engineering Division, University of Glasgow, Glasgow, G12 8QQ, United Kingdom}

\author{Martin P. Weides}
\affiliation{James Watt School of Engineering, Electronics \& Nanoscale Engineering Division, University of Glasgow, Glasgow, G12 8QQ, United Kingdom}

\author{Rair Macêdo}
\email{Rair.Macedo@glasgow.ac.uk}
\affiliation{James Watt School of Engineering, Electronics \& Nanoscale Engineering Division, University of Glasgow, Glasgow, G12 8QQ, United Kingdom}

\date{\today}

\begin{abstract}
    Evidence of non-hermitian behavior has been recently demonstrated in cavity magnonics, including the emergence of mode level attraction and exceptional points in spectroscopic measurements. This work demonstrates experimental evidence of time-domain dynamics of magnon-photon systems that are coupled through a long-range interaction (i.e. remote coupling) exhibiting level attraction mediated by an auxiliary mode. 
    We directly observe the temporal evolution of dissipatively coupled cavity-magnon modes, where heavily damped transmission line modes mediate the interaction. 
    Our frequency-domain measurements confirm the predicted level attraction, while time-domain ring-down measurements reveal the characteristic signatures of dissipative coupling dynamics. 
    Our approach offers in situ tunability over the dissipative coupling strength, including complete suppression, without requiring physical modifications to the experimental setup, providing a versatile platform for exploring tunable, non-Hermitian physics. 
    \end{abstract}

\maketitle


In cavity magnonics, magnons---the quanta of collective excitations in spin systems---are studied as the primary intermediary between a range of other excitations; from optical and microwave photons, to phonons and qubits \cite{lachance-quirion_hybrid_2019,jiang2023integrating}. 
Remote coupling in cavity-magnon systems, where a magnetic sample is positioned outside the cavity field \cite{macedo_CMPreview}, has recently emerged as a promising route for novel developments in these architectures harnessing the unique properties of magnetic systems \cite{yang_anomalous_2024, kim_microwave_2025, van_loo_photon-mediated_2013,xiang_hybrid_2013, clerk_hybrid_2020}.  
Unlike directly coupled systems where the magnetic element resides within the cavity mode volume, remote coupling enables flexible spatial arrangements whilst maintaining magnon-photon interactions over macroscopic distances~\cite{li_coherent_2022,rao_meterscale_2023, rao_braiding_2024, xiong_magnon-photon_2024}. Distributed magnonic nodes could serve as quantum memories or transducers, interfacing through shared modes that act as a bus for information transfer and manipulation.

The frequency-domain properties of remote cavity-magnonic systems have recently been well characterized, revealing coupling mechanisms fundamentally different from their directly coupled counterparts. The indirect magnon-photon interactions of remote coupling have been shown to exhibit conventional level repulsion, that is, the coupled modes exhibit avoided crossings and coherent energy exchange. On the other hand, dissipative coupling mechanisms in such geometries have been shown to give rise to level attraction---where a complex coupling leads modes to coalesce rather than repel \cite{li_coherent_2022,rao_meterscale_2023,yang_anomalous_2024}.

Before the development of remotely coupled cavity–magnon systems, level attraction spectra had been observed in a number of experimental platforms and attributed to a variety of mechanisms \cite{wang2020dissipative}. Its observation in cavity magnonics has opened a rich landscape \cite{hurst2022non} for experimental exploration of non-Hermitian physics, including the emergence of reflectionless or topologically braided states~\cite{rao_braiding_2024} and the use of topological operations around exceptional points to coherently manipulate strongly coupled magnon–photon dynamics \cite{lambert_coherent_2025}.

Such non-hermitian physics has most commonly been engineered through the realization of systems with balanced loss and gain \cite{el2018non, zhang_observation_2017}, as these can satisfy the $\mathcal{PT}$-symmetry required to still exhibit real eigenvalues \cite{Bender98}. 
In cavity-magnonics, effective dissipative coupling between the cavity and magnon subsystems can give rise to level attraction and \textit{anti}-$\mathcal{PT}$ symmetric Hamiltonians through two distinct mechanisms: \emph{(i)} coupling via a common travelling wave continuum, where both modes dissipate into the shared bath, and \emph{(ii)} coupling via a heavily damped auxiliary mode \cite{MetelmannClerk, Yu2019, Zhao2020, Yang2020}.

Although both mechanisms lead to the same general features of mode coalescence observed in spectroscopic measurements, many of the unique properties being explored in non-Hermitian physics---such as chiral mode switching \cite{doppler2016dynamically,zhang2019dynamically}, energy transfer \cite{xu2016topological}, and potential sensing applications \cite{Scullly2021}---are inherently dynamic processes that may exhibit distinct signatures in the time domain. 
While recent work has investigated the temporal evolution of modes indirectly coupled through traveling waves ~\cite{lu_temporal_2025, lambert_coherent_2025}, the auxiliary-mode mechanism---where a transmission line can act as a heavily damped intermediary between cavity and magnon modes \cite{Yu2019}---may exhibit temporal signatures that differ from those of travelling-wave coupling, which remain unexplored. 
At the same time, the in-situ control of dissipative coupling has been a longstanding goal in cavity magnonics since these experimental signatures were first observed. 
In particular, phase control realized in multi-drive architectures \cite{wolz_introducing_2020, joseph_role_2024} has been attractive because it does not require active gain to engineer tunable non-Hermitian Hamiltonians \cite{grigoryan_synchronized_2018, boventer_steering_2019,boventer_control_2020}, though it was recently shown \cite{gardin2025level, smith_exceptional_2025} these types of multi-drive systems are distinct from systems realizing level attraction through a genuine dissipative coupling. 

Here, we demonstrate phase-controlled tunable dissipative coupling between spatially separated cavity-magnon modes, where heavily damped transmission line modes act as the mediating auxiliary.
By adjusting the transmission line detuning, we introduce continuous control from strong level attraction to complete decoupling---all without physically repositioning any components. 
Our frequency-domain measurements confirm the predicted level attraction, while time-domain ring-down measurements reveal how this dissipative coupling manifests in the system's temporal evolution. 
The theoretical framework, based on the auxiliary mode mechanism, accurately captures both the frequency-domain spectra and the observed time-domain dynamics.
Taken together, such understanding and control of these temporal dynamics are key steps toward integration of cavity magnonics into devices for applications such as quantum information processing, where precise control over the time evolution of quantum states is essential.

\begin{figure}[htb]
\includegraphics[width=\linewidth]{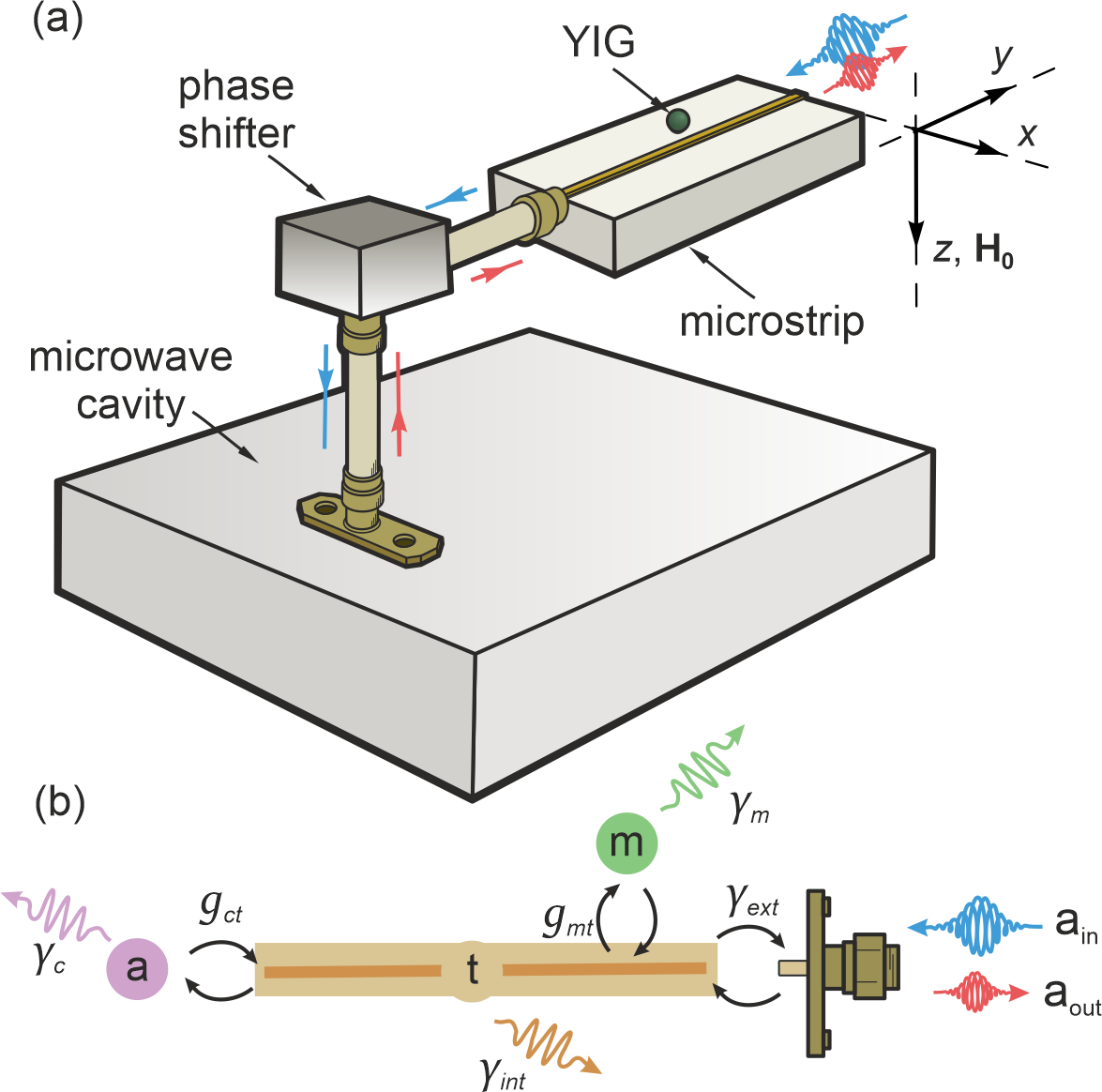}
\caption{(a) Experimental diagram of the remote cavity–magnon coupling system. A YIG sphere on a microstrip transmission line is coupled to a three-dimensional microwave cavity via a variable phase shifter. (b) Schematic model showing the cavity, transmission-line, and magnon modes with their associated coupling and dissipation channels.}
\label{fig:remoteCouplingSetUp} 
\end{figure}

\emph{The system}---Shown in Fig.~\ref{fig:remoteCouplingSetUp}(a), our system consists of a microwave signal propagating along a microstrip transmission line, with a YIG sphere positioned on top. This microstrip is connected through a variable phase shifter to a three-dimensional microwave cavity resonator, and the reflection from this path is measured. The YIG sphere is subject to an external static magnetic field $\mathbf{H}_{0}$, allowing tuning of the magnon resonance frequency. The cavity photons and magnons in the YIG are not directly coupled; instead, they interact via the electromagnetic fields of the transmission line.
The system dynamics can be well-approximated by a linear three-mode model with a Hamiltonian of the form

\begin{equation}
\begin{split}
H = \hbar \omega_c a^\dagger a + \hbar \omega_m m^\dagger m + \hbar \omega_{t} t^\dagger t \\[8pt]
+ \hbar g_{mt}(m^\dagger t + m t^\dagger) + \hbar g_{ct}(a^\dagger t + a t^\dagger).
\end{split}
\end{equation}
Here, $a^\dagger$ and $a$ ($m^\dagger$ and $m$) denote the creation and annihilation operators for cavity photons (magnons). 
The transmission line is represented by a single effective bus mode with frequency $\omega_t$, linewidth $\gamma_t$, and $t^\dagger$ and $t$ denoting its creation and annihilation operators. 
The frequencies $\omega_c$ and $\omega_m$ represent the cavity and magnon resonance, with $\omega_m$ tunable via $\mathbf{H}_0$ by $\omega_m =\gamma\mu_0H_0$. 
The coupling strengths $g_{mt}$ and $g_{ct}$ describe the YIG-transmission line coupling (controllable via phase shifter) and cavity-transmission line coupling. 
The phase shifter provides independent control over both the magnon coupling strength to the transmission line $g_{mt}$ and the effective frequency $\omega_{t}$ (see Supplemental Material).

Physically, the super-damped intermediary acts like a broadband bath, where we have $\gamma_t \gg \gamma_c, \gamma_m$ ($\gamma_c $ and $ \gamma_m$ being the dissipation rates of the cavity and magnon, respectively) and the total dissipation rate for the transmission-line mode  
$\gamma_{t} = \gamma_{\mathrm{int}} + \gamma_{\mathrm{ext}}$, where $\gamma_{\mathrm{int}}$ accounts for intrinsic (internal) losses and $\gamma_{\mathrm{ext}}$ describes coupling to the external port through which the input field $a_{\mathrm{in}}$ and output field $a_{\mathrm{out}}$ are defined [as summarized in Fig.~\ref{fig:remoteCouplingSetUp}(b)]. 
To model decay and measurements, the equations of motion become:
\begin{align}
\dot{a} &= -i\omega_c a - ig_{ct} t - \frac{\gamma_c}{2} a \label{eq:cavity} \\
\dot{m} &= -i\omega_m m - ig_{mt} t - \frac{\gamma_m}{2} m \label{eq:magnon} \\
\dot{t} &= -i\omega_{t} t - ig_{ct} a - ig_{mt} m - \frac{\gamma_{t}}{2} t + \sqrt{\gamma_{\mathrm{ext}}}\, a_{\mathrm{in}}, \label{eq:transmission}
\end{align}
To obtain the frequency-domain response, namely the reflection scattering parameter $S_{11}(\omega)$, we employ input-output theory (including resulting solutions for the bus-mode). This gives
\begin{equation}
S_{11}(\omega) = 1 - \frac{\gamma_{\mathrm{ext}}}{i\Delta_t + \frac{\gamma_{t}}{2} + \frac{g_{ct}^2}{i\Delta_c + \frac{\gamma_c}{2}} + \frac{g_{mt}^2}{i\Delta_m + \frac{\gamma_m}{2}}}
\label{eq:S11}
\end{equation}
where $\Delta_c = \omega_c - \omega$, $\Delta_m = \omega_m - \omega$, and $\Delta_t = \omega_t - \omega$ are the detunings of the three modes from the drive.

\begin{figure}[htb]
\includegraphics[width=\linewidth]{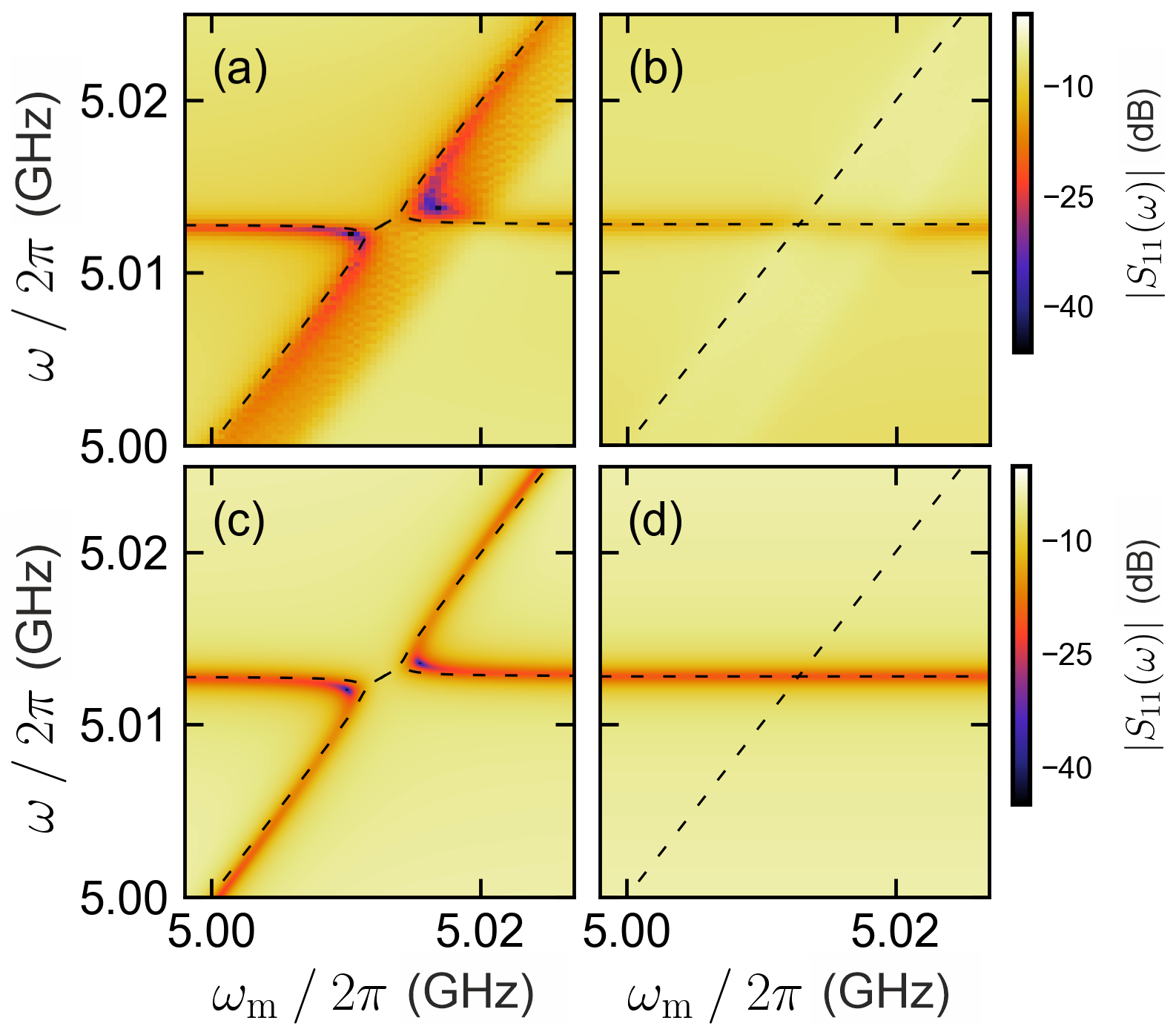}
\caption{Frequency spectrum of the remote coupling system showing two distinct regimes. (a) Level attraction between cavity and magnon modes arising from dissipative coupling when $\omega_t = \omega_c$.  (b) Complete magnon decoupling, $g_{mt} = 0$ and $\omega_t \neq \omega_c$. (c,d) Theoretical predictions using Eq.~\ref{eq:S11} for panels (a) and (b), respectively. Dashed lines in all panels denote the hybridised eigenfrequencies $\omega_\pm$, calculated from 
Eq.~\ref{eq:analytical_eigenfreqs_final}} 
\label{fig:frequencySpectrum}
\end{figure}

The resulting experimental and theoretical spectra for our system are shown in Fig.~\ref{fig:frequencySpectrum} summarizing two distinct coupling regimes: dissipative coupling with the distinct mode level attraction spectra [Fig.~\ref{fig:frequencySpectrum}(a) and (c)] and complete mode decoupling [Fig.~\ref{fig:frequencySpectrum}(b) and (d)].
These are accessed through controlling the phase relationship between forward and backward traveling waves (via the phase shifter).

Before discussing these any further, it is useful to investigate the effective interaction between the cavity and the magnon by eliminating the transmission-line mode 
which yield a simplified two-mode model:

\begin{equation}
\label{eq:am-matrix}
\begin{bmatrix}
z_c(\omega) & g_{\mathrm{eff}}(\omega) \\[6pt]
g_{\mathrm{eff}}(\omega) & z_m(\omega)
\end{bmatrix}
\begin{bmatrix} a \\[0.5em] m \end{bmatrix}
=
\frac{\sqrt{\gamma_t}}{i\Delta_t + \frac{\gamma_t}{2}}
\begin{bmatrix} g_{ct} \\[0.5em] g_{mt} \end{bmatrix}
a_{\mathrm{in}}^{\,0}.
\end{equation}
with
\begin{equation}
z_n(\omega)
= -\Delta_n + i\frac{\gamma_n}{2}
+ \frac{g_{nt}^2}{\Delta_t - i\frac{\gamma_t}{2}},
\quad n\in\{c,m\}.
\end{equation}
Here, the cavity and magnon modes do not interact directly but couple through the highly damped transmission line as an intermediary mode. This creates an \textit{effective coupling} between the cavity and magnon given by:
\begin{equation}
g_{\mathrm{eff}}(\omega)
= \frac{g_{ct} g_{mt}}{\Delta_t - i\frac{\gamma_t}{2}}.
\label{eq:g_eff}
\end{equation}

The term $g_{\mathrm{eff}}(\omega)$ encapsulates how the transmission line mediates both energy exchange and correlated dissipation between the two modes. 
The behaviour observed in Fig.~\ref{fig:frequencySpectrum} can be understood within the framework of dissipative coupling mediated by an auxiliary mode \cite{Yu2019, Zhao2020, MetelmannClerk}. 
The effective coupling between the cavity and magnon given in Eq.~\ref{eq:g_eff} is inherently complex and frequency-dependent. This contrasts with direct coherent coupling (which would be purely real) and reveals the dissipative nature of the interaction. 
\begin{equation}
g_{\mathrm{eff}}(\omega) = g_{\mathrm{coh}}(\omega) + i\Gamma_{\mathrm{diss}}(\omega),
\end{equation}
where
\begin{align}
g_{\mathrm{coh}}(\omega) &= \frac{g_{ct} g_{mt} \Delta_t}{\Delta_t^2 + (\gamma_t/2)^2}, \label{eq:coherent_part}\\
\Gamma_{\mathrm{diss}}(\omega) &= \frac{g_{ct} g_{mt} (\gamma_t/2)}{\Delta_t^2 + (\gamma_t/2)^2}. \label{eq:dissipative_part}
\end{align}

\emph{Regime 1: Dissipatively coupled modes}---When the transmission line resonance is matched to the cavity frequency ($\omega_t = \omega_c$) and the YIG sphere is positioned at a field antinode ($g_{mt} \neq 0$), we observe level attraction between the cavity and magnon modes [Fig.~\ref{fig:frequencySpectrum}(a)]. Thus, demonstrating that even though the cavity and magnon are spatially separated and not directly coupled, the transmission line successfully mediates their interaction. The auxiliary mode acts as a common reservoir for the cavity and magnon, mediating dissipative interactions between them, as described by the effective coupling in Eq.~\ref{eq:g_eff}.

The theoretical prediction shows excellent agreement with experimental data [Fig.~\ref{fig:frequencySpectrum}(c)]. 
In the strongly damped limit where $\gamma_t \gg |\Delta_t|$, the effective coupling becomes predominantly dissipative:
\begin{equation}
g_{\mathrm{eff}} \approx i\frac{2g_{ct} g_{mt}}{\gamma_t}.
\label{eq:markovian_limit}
\end{equation}
This purely imaginary coupling reaches its maximum magnitude when the transmission-line mode is on resonance ($\Delta_t = 0$), at which point the coherent component $g_{\mathrm{coh}}$ vanishes [Eq.~\ref{eq:coherent_part}] while the dissipative component $\Gamma_{\mathrm{diss}}$ is maximised [Eq.~\ref{eq:dissipative_part}]. Experimentally, we achieve this resonance condition by tuning the phase shifter to bring the transmission line frequency into degeneracy with the cavity mode, as characterised in the Supplemental Note B [Fig.~\ref{fig:tranmissionLineTuning}]. 

The coupled cavity–magnon dynamics can be expressed directly as two linear equations for the complex mode amplitudes $a$ and $m$. In the strongly damped limit $\gamma_t/2 \gg |\Delta_t|$, the transmission-line mode can be eliminated and the hybridised eigenmodes of the coupled cavity–magnon system obey:
\begin{equation}
\omega_{\pm} = \bar{\omega} - i\bar{\gamma}
\pm \sqrt{\left(\frac{\Delta}{2}
                - i\frac{\delta\gamma}{4}\right)^2
          - \Gamma^2},
\label{eq:analytical_eigenfreqs_final}
\end{equation}
where $\bar{\omega} = (\omega_c + \omega_m)/2$ is the average frequency, $\bar{\gamma} = (\gamma_c' + \gamma_m')/4$ is the average damping, $\Delta = \omega_c - \omega_m$ is the cavity--magnon detuning, and $\delta\gamma = \gamma_c' - \gamma_m'$ is the damping difference. These eigenfrequencies are shown as dashed lines in Fig.~\ref{fig:frequencySpectrum}.
The nature of the eigenmodes depends on the argument of the square root in Eq.~\eqref{eq:analytical_eigenfreqs_final}. When $\Gamma^2 < [(\Delta/2)^2 + (\delta\gamma/4)^2]$, the square-root term is predominantly real, yielding two distinct hybrid modes with different resonance frequencies. 
At the exceptional point, $\Gamma^2 = [(\Delta/2)^2 + (\delta\gamma/4)^2]$, the square root vanishes and the eigenfrequencies coalesce
. In the opposite, strong-coupling regime where $\Gamma^2 > [(\Delta/2)^2 + (\delta\gamma/4)^2]$, the square-root term becomes predominantly imaginary, leading to identical resonance frequencies. 


\emph{Regime 2: Mode decoupling}---By controlling the phase between backwards and forward traveling waves, we change two parameters. First, we detune the tranmission line mode: and where $|\Delta_t| \gg \gamma_t$, the coherent component dominates:
\begin{equation}
g_{\mathrm{eff}} \approx \frac{g_{ct} g_{mt}}{\Delta_t},
\end{equation}
however, the large $\Delta_t$ in the denominator makes $g_{\mathrm{eff}} \rightarrow 0$.

Secondly, we tune the standing wave pattern to position the YIG sphere at a field node, effectively eliminating magnon-transmission line coupling ($g_{mt} \rightarrow 0$). This results in complete decoupling of the magnon from the system [Fig.~\ref{fig:frequencySpectrum}(b)], where only the cavity response remains visible. This demonstrates precise control over coupling strength without mechanical movement—a key advantage of the phase-controlled approach. The theoretical model [Fig.~\ref{fig:frequencySpectrum}(d)] confirms that setting $g_{mt}$ = 0 in Eq.~\ref{eq:S11} reproduces the experimental observation of an isolated cavity resonance. 
This is consistent with Eq.~\ref{eq:g_eff}: in the absence of the individual 
coupling strengths $g_{ct}$ and $g_{mt}$, we have  $g_{\mathrm{eff}}=0$, meaning there is no effective interaction between the cavity and magnon modes, as expected for spatially separated systems with no direct coupling pathway.
Essentially the phase shifter control directly eliminates both coupling components in Eqs.~\ref{eq:coherent_part}--\ref{eq:dissipative_part}.


\emph{Time-Domain Dynamics}---To probe the temporal evolution of the coupled system, we performed ringdown measurements using short pulse excitation. A 16~ns square pulse at the fixed frequency ($\omega_c$) was applied to the input port, while the magnon frequency was tuned via $\mathbf{H_0}$. 
The reflected signal was measured by monitoring the microstrip output voltage with a sampling interval of 1.63~ns. 

\begin{figure}[htbp]
\includegraphics[width=\linewidth]{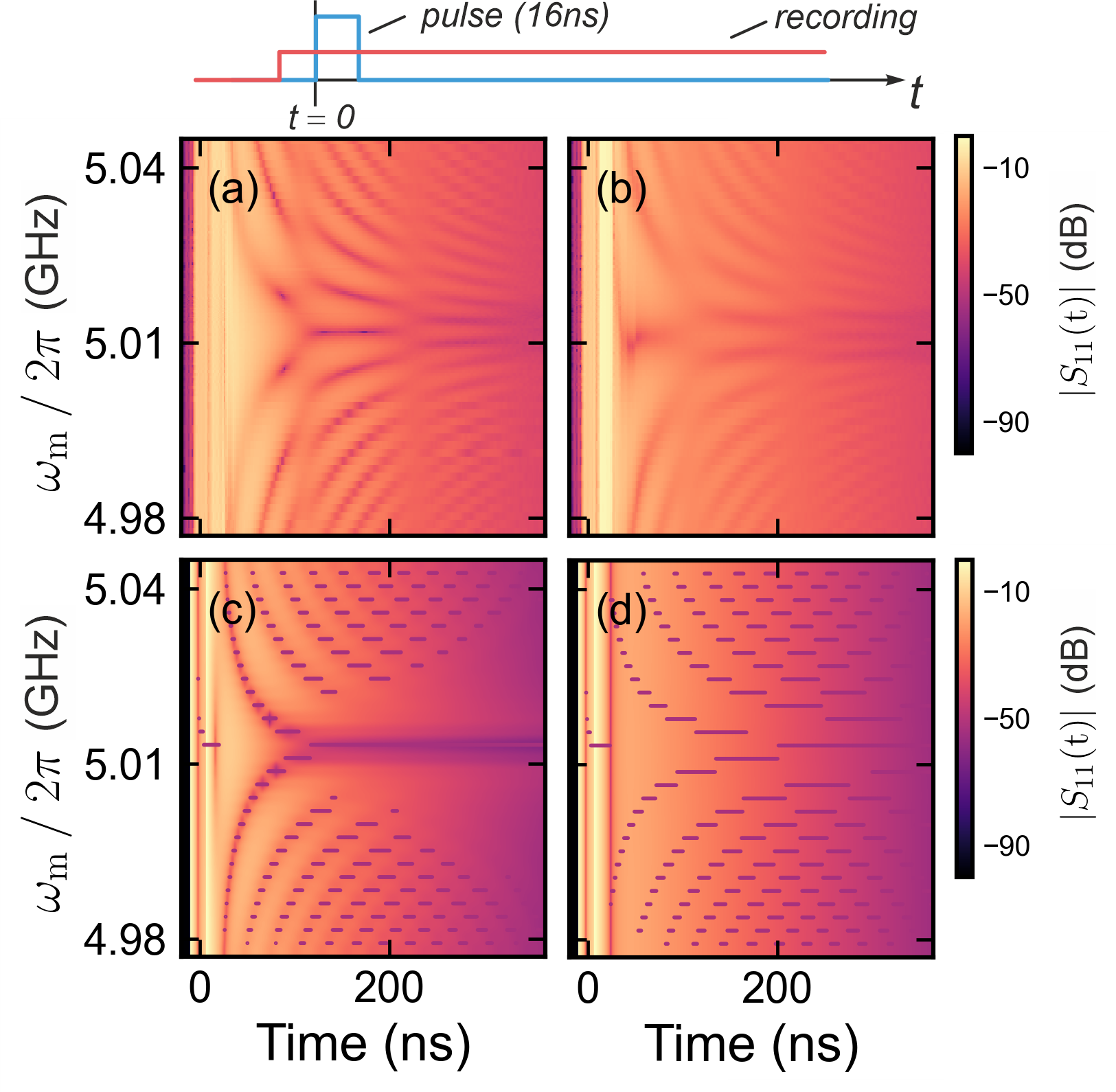}
\caption{Time-domain response of the remote coupling system. (a) Dissipative coupling regime showing Ramsey-like interference as the system is tuned through level attraction. (b) Decoupled regime. (c,d) Theoretical calculations using Eqs.~\ref{eq:cavity} -- \ref{eq:transmission} corresponding to (a) and (b), respectively.} 
\label{fig:levelAttractionTimeSpectrum} 
\end{figure}

Fig.~\ref{fig:levelAttractionTimeSpectrum} shows the time-domain dynamics for both coupling regimes as the magnon frequency is tuned through resonance. The dissipative coupling regime [Fig.~\ref{fig:levelAttractionTimeSpectrum}(a)] is achieved at a phase shifter setting where $\omega_t = \omega_c$, while the decoupled regime [Fig.~\ref{fig:levelAttractionTimeSpectrum}(b)] is achieved at a different phase shifter setting where $\omega_t \neq \omega_c$ and $g_{mt} \approx 0$. Theoretical predictions [Figs.~\ref{fig:levelAttractionTimeSpectrum}(c) and~\ref{fig:levelAttractionTimeSpectrum}(d)] show excellent agreement with experimental data.

The time-domain response in the dissipative coupling regime exhibits significantly different behavior compared to conventional level repulsion systems, showing a Ramsey-like interference pattern rather than the quasi-Rabi oscillations typically observed in strongly coupled systems. 
Far from resonance, the ringdown signal exhibits characteristic beating behavior. As the bias field brings the system closer to level attraction ($\omega_m \approx \omega_c = 5.012$~GHz), the beating pattern diminishes and the two oscillation frequencies converge, with enhanced dissipation (rapid energy absorption) occurring at exact resonance. This temporal signature directly reflects the coalescence of the two hybridized eigenmodes as the system approaches the exceptional point, where the non-Hermitian Hamiltonian transitions from having two distinct complex eigenvalues to a single degenerate eigenvalue.

In contrast, the decoupled regime [Fig.~\ref{fig:levelAttractionTimeSpectrum}(b)] shows significantly reduced dynamics. While some weak beating is still observable far from resonance due to residual magnon-transmission line coupling ($g_{mt}$ close to but not exactly zero), the interaction strength is greatly diminished compared to the dissipative coupling regime. Since $g_{\mathrm{eff}} \propto g_{mt}$ [Eq.~\ref{eq:g_eff}], minimizing $g_{mt}$ directly suppresses the effective cavity-magnon interaction. The mode convergence on resonance occurs much more gradually, and the temporal response is essentially identical to that of an isolated cavity. 

Theoretical calculations from the coupled equations of motion [Eqs.~\ref{eq:cavity}--\ref{eq:transmission}] reproduce the essential features of both regimes [Figs.~\ref{fig:levelAttractionTimeSpectrum}(c,d)], capturing both Ramsey-like interference and weak dynamics due to a vanising $g_{mt}$. 
Note that the parameters used here differ from those extracted from frequency-domain measurements [Fig.~\ref{fig:frequencySpectrum}], as our theoretical model uses a simplified single transmission line mode for clarity. Despite this simplification, the qualitative agreement validates the underlying physics. The complete model incorporating both transmission line modes is presented in Supplemental Note~F.


This demonstrates the tunability of auxiliary mode-mediated coupling. We can vary the effective coupling strength without repositioning any components---dynamically controlling interactions between the magnon and the cavity. 
The beating patterns observed off-resonance in the coupled regime indicate longer-lived hybrid states, providing tunable access between rapid dissipation at resonance and extended coherence times away from resonance. 







\emph{Discussion and Conclusions}---We have demonstrated how an auxiliary mode can lead to  dissipative coupling and ultimately, long-range coupled cavity-magnon system, providing evidence of the time-domain behavior and insight into the nature of this coupling mechanism. 
The auxiliary mode mechanism realized in our system differs fundamentally from that recently studied in Ref.~\cite{lu_temporal_2025}, where both resonators couple to the same traveling wave bath.

In travelling wave coupling, the dissipative interaction arises as $\Gamma = \sqrt{\kappa_m \kappa_c}$ determined by the extrinsic system-bath coupling rates. In contrast, our auxiliary mode mechanism yields an effective dissipative coupling $\Gamma_{\mathrm{diss}} = 2g_{ct}g_{mt}/\gamma_t$ [Eq.~\ref{eq:markovian_limit}] that depends on the product of the individual mode-auxiliary couplings and is inversely proportional to the auxiliary mode damping. 
This fundamental difference provides distinct parameter spaces for engineering dissipative interactions. 
In particular, we note the auxiliary-mode damping $\gamma_t$ can be influenced by impedance mismatch at the transmission line terminations, which controls reflections and therefore the effective mode characteristics. A comparable leaky waveguide mode appears in the COMSOL simulations of Ref.~\cite{Yu2019}; while the physical origin of that mode was not specified, impedance mismatch at waveguide boundaries offers a plausible mechanism consistent with our analysis. 
A key advantage of our implementation is the ability to continuously tune the coupling strength through control of the phase between backward and forward traveling waves in the cable path, thus requiring no mechanical repositioning. 
This tunability arises from the dependence of $g_{mt}$ on the spatial configuration, which can be effectively modified through phase control of the transmission line.



Under appropriate conditions, systems with purely dissipative coupling can exhibit anti-$\mathcal{PT}$ symmetry and exceptional points~\cite{Yang2020, Zhao2020}. 
Following Ref.~\cite{Zhao2020}, our system would approach anti-$\mathcal{PT}$ symmetry if: (i) the auxiliary mode damping dominates detunings: $\gamma_t \gg |\omega_c - \omega_t|$, $|\omega_m - \omega_t|$ (satisfied in our strong coupling regime), (ii) the renormalized dissipation rates of the modes are equal: $\gamma_c' = \gamma_m'$, and (iii) the coupling is purely dissipative with negligible coherent component (see Supplemental Note E). Whilst our system naturally satisfies condition (i), achieving equal damping (ii) can often be challenging due to mismatch between cavity and magnon mode dissipation, though relevant means of extrinsic damping control \cite{uno2025tunable} have recently been explored. In our system however, $\gamma_m'$ is tunable through the phase dependence of $g_{mt}$.



Recent work has shown that dissipatively coupled systems can support long-lived hybrid modes and entanglement between subsystems~\cite{nair_ultralow_2021}, suggesting that our platform could enable enhanced coherence times relevant for magnon-based quantum information processing. 
Furthermore, anti-$\mathcal{PT}$-symmetric systems have been shown to exhibit unique quantum correlation dynamics \cite{NoriAntiPT}, including periodic entanglement oscillations near exceptional points. Our ability to tune through exceptional points via phase control could enable exploration of these phenomena in magnonic systems, where the controllable transition between coupling regimes may offer new pathways for entanglement generation and manipulation in open quantum systems.



\section*{Data availability}

All experimental data presented in this study is available in the Enlighten database at
(DOI to be generated by the University of Glasgow once final article is accepted).

\begin{acknowledgments}
This work was supported by the Science and Technology Facilities Council (STFC) under grant number UKRI/ST/B000600/1.
M. A. Smith acknowledges funding from the Engineering and Physical Sciences Research Council
(EPSRC) under grant number EP/S023321/1.
\end{acknowledgments}

\section*{Author Contributions}

A.J. conceived the experiment, performed measurements with input from R.M. and M.P.W., carried out data analysis with contributions from M.A.S., and derived the theoretical model with support from M.A.S. A.J. wrote the manuscript with input and discussions from all coauthors. M.P.W. and R.M. supervised the project.

\section*{Competing Interests}

The authors declare no competing interests.

\bibliographystyle{apsrev4-2}
\bibliography{references}

\newpage
\clearpage


\begingroup

\makeatletter
\fontsize{12}{20}\selectfont
\setlength{\textwidth}{6.5in}
\setlength{\textheight}{9.0in}
\setlength{\oddsidemargin}{0in}
\setlength{\evensidemargin}{0in}
\setlength{\topmargin}{-0.625in}
\setlength{\headheight}{12pt}
\setlength{\headsep}{25pt}
\setlength{\footskip}{30pt}
\setlength{\parindent}{0.5cm}
\setlength{\parskip}{0pt}
\setcounter{page}{1}      
\def\ps@plain{%
  \let\@mkboth\@gobbletwo
  \def\@oddhead{}\def\@evenhead{}%
  \def\@oddfoot{\hfill\fontsize{12}{25}\selectfont\thepage\hfill}%
  \let\@evenfoot\@oddfoot}
\pagestyle{plain}

\newcommand{\suppsectionsize}{\fontsize{11}{13}\selectfont}     
\newcommand{\suppsubsectionsize}{\fontsize{11}{12.5}\selectfont}
\newcommand{\suppsubsubsectionsize}{\fontsize{11}{12.5}\selectfont}

\renewcommand\section{\@startsection{section}{1}{1.25em}%
  {-3.2ex plus -1ex minus -.2ex}%
  {2.5ex plus .3ex}%
  {\raggedright\normalfont\bfseries\suppsectionsize}}

\renewcommand\subsection{\@startsection{subsection}{2}{1.25em}%
  {3.5ex plus 0.6ex}%
  {2.4ex plus .3ex}%
  {\raggedright\normalfont\bfseries\suppsubsectionsize}}

\renewcommand\subsubsection{\@startsection{subsubsection}{3}{1.25em}%
  {-2.4ex plus -0.8ex minus -.2ex}%
  {2.4ex plus .3ex}%
  {\raggedright\normalfont\itshape\suppsubsubsectionsize}}

\def\@afterheading{%
  \@nobreaktrue
  \everypar{%
    \if@nobreak
      \@nobreakfalse
      \clubpenalty\@M
      \everypar{}%
    \else
      \everypar{}%
    \fi}}

\long\def\@makecaption#1#2{%
  \vskip 0.5em
  \fontsize{11.25}{20}\selectfont
  #1.\; #2\par
}

\makeatother
\onecolumngrid

\renewcommand{\theequation}{S\arabic{equation}}
\renewcommand{\thetable}{S\arabic{table}}
\renewcommand{\thefigure}{S\arabic{figure}}
\setcounter{equation}{0}
\setcounter{table}{0}
\setcounter{figure}{0}
\setcounter{page}{0}











\clearpage
\begin{center}

{\fontsize{15}{20}\selectfont\bfseries
Supplemental Material for:\\[0.75em]
Dynamic Modulation of Long-Range Photon–Magnon Coupling}

\vspace{2.0em}

{\fontsize{12.5}{15}\selectfont
Alban Joseph,$^{1,*}$ 
Mawgan A. Smith,$^{1}$ 
Martin Weides,$^{1}$ 
Rair Macêdo$^{1}$}

\vspace{1.25em}

{\fontsize{11}{20}\selectfont\itshape
$^{1}$James Watt School of Engineering, Electronics \& Nanoscale Engineering Division,\\
University of Glasgow, Glasgow, G12 8QQ, United Kingdom}

\vspace{0.75em}

{\fontsize{11}{20}\selectfont $^{*}$a.joseph.2@research.gla.ac.uk}

\end{center}

\vspace{2.5em}


\cleardoublepage
\section*{Supplementary Note A: Phase Shifter Characterization}
\label{sec:phase_shifter_char}

To establish the relationship between phase shifter settings and propagating signals, we first characterized the phase shifter response with the cavity removed from the setup. Figure~\ref{fig:phaseShifter} shows the reflection coefficient phase measured at 5.012~GHz as a function of phase shifter setting $\varphi$.

\begin{figure}[htb]
\centering
\includegraphics[width=0.8\linewidth]{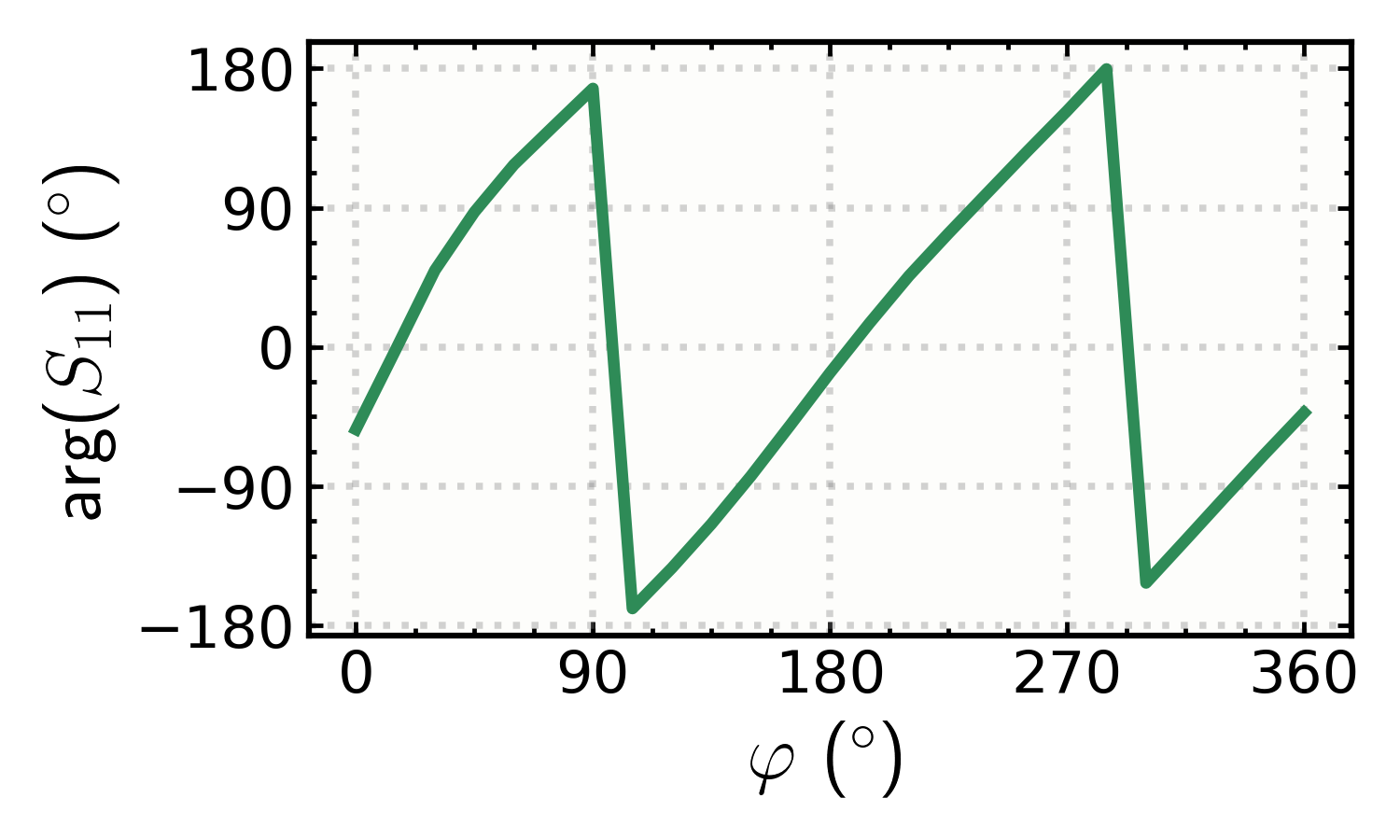}
\caption{Reflection coefficient phase at frequency = 5.012GHz for different phase shifter settings $\varphi$.}
\label{fig:phaseShifter}
\end{figure}

Since the signal traverses the phase shifter twice (forward and reflected paths), the effective phase shift is doubled compared to the control setting. This calibration measurement established the baseline relationship between phase shifter control value and actual phase shift in the signal.

\newpage

\section*{Supplementary Note B: Transmission Line Resonator Characterization}
\label{sec:tl_char}
Although the ideal transmission line should simply constitute of a continuum of propagating TEM modes like coaxial cables, the boundary conditions imposed by the cavity termination, finite line length, and any impedance mismatch can create quasi-discrete standing wave resonances. Changing the phase shifter setting alters the total electrical length between the transmission line path and the cavity termination, which modifies the resonance conditions for standing wave modes, effectively shifting the positions of the nodes and antinodes along the transmission line. This phase control thus provides a mechanism to continuously tune the transmission line detuning ($\Delta_t$) without mechanical movement.

We first characterized the bare transmission line resonator properties with the cavity disconnected from the setup. Figure~\ref{fig:tlDamping} shows a representative reflection spectrum at $\varphi = 180^\circ$, revealing the highly damped bus modes. Using circle fit analysis on the complex $S_{11}$ data for one of these modes, we extracted the loaded quality factor $Q_L = 98.6$, internal quality factor $Q_i = 547.4$, and coupling quality factor $Q_c = 120.3$, corresponding to damping rates $\gamma_t/2\pi = 50.8$~MHz, $\gamma_{\mathrm{int}}/2\pi = 9.16$~MHz, and $\gamma_{\mathrm{ext}}/2\pi = 41.7$~MHz respectively. 

\begin{figure}[htb]
\centering
\includegraphics[width=0.8\linewidth]{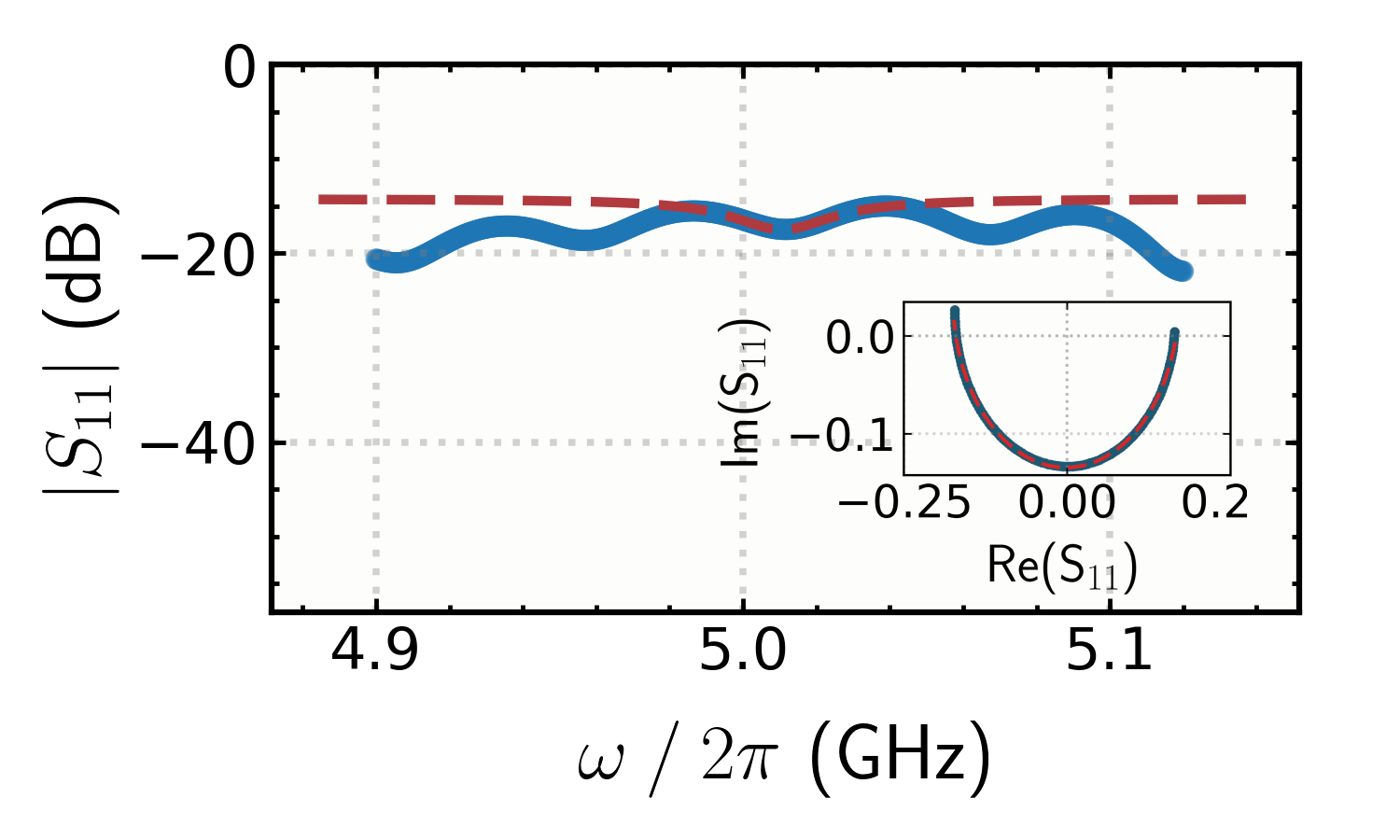}
\caption{Reflection spectrum $|S_{11}|$ of the transmission line resonator with circle fit (inset) yielding $\gamma_t/2\pi = 50.8$~MHz, $\gamma_{\mathrm{int}}/2\pi = 9.16$~MHz, and $\gamma_{\mathrm{ext}}/2\pi = 41.7$~MHz.}
\label{fig:tlDamping} 
\end{figure}

Approximately 82\% of the resonator energy escapes through coupling to the measurement port ($\gamma_{\mathrm{ext}}/\gamma_t \approx 0.82$), with only a small fraction dissipated internally, indicating that the majority of the energy leaves the resonator through the port rather than being lost internally.

With the cavity connected to the transmission line and with the YIG bias field turned off (no magnon resonance), we performed reflection measurements as a function of both frequency and phase shifter setting (Figure~\ref{fig:tranmissionLineTuning}a). This baseline measurement reveals how transmission line resonances shift with phase changes (marked with dashed lines) and identifies phase settings where the transmission line mode becomes degenerate with the cavity mode. The transmission line resonance exhibits an approximately linear dependence on phase shifter setting, with a tuning rate of $\sim$0.3~MHz per degree:
\begin{align}
\omega_{t1}(\varphi) &= 2\pi \times (0.319\varphi + 4981.6)~\text{MHz}, \label{eq:supomega_t1} \\
\omega_{t2}(\varphi) &= 2\pi \times (0.308\varphi + 4928.9)~\text{MHz}. \label{eq:supomega_t2}
\end{align}
where $\varphi$ is the phase shifter setting in degrees.

\begin{figure}[hbt]
\centering
\includegraphics[width=0.95\linewidth]{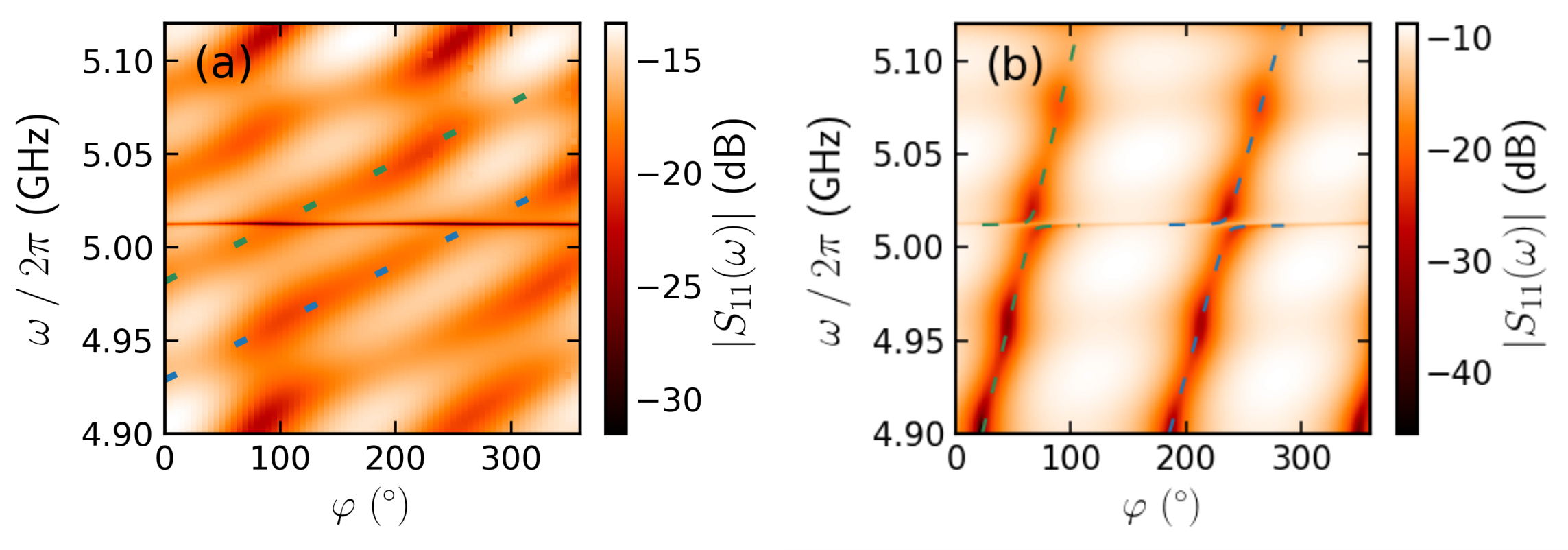}
\caption{Reflection coefficient vs frequency for different phase shifter settings, showing (a) highly damped microstrip transmission line tuning; dashed lines indicate the phase-dependent shift of the mode resonance, and (b) higher-Q coplanar waveguide resonator tuning; revealing avoided crossing used to estimate the cavity--transmission line coupling strength, $g_{ct}$.} 
\label{fig:tranmissionLineTuning}
\end{figure} 

The transmission line in this configuration acts as a highly damped resonator with $\gamma_t \gg \gamma_c$. To estimate the cavity--transmission line coupling strength $g_{ct}$, we substituted the microstrip transmission line with a higher-$Q$ coplanar waveguide (CPW) resonator. This substitution enabled observation of clear anti-crossing behavior between the cavity and transmission line modes (Figure~\ref{fig:tranmissionLineTuning}b), with an avoided crossing gap of approximately 9.8~MHz at degeneracy. This yields a coupling strength of $g_{ct}/2\pi \approx$ 4.9~MHz. 

The coupling strength extracted from this measurement provides the $g_{ct}$ parameter required for the theoretical model. With the high-$Q$ resonator, normal-mode splitting becomes observable since the condition $g_{ct} \gtrsim (\gamma_c + \gamma_t)/8$ can be satisfied, whereas the original highly damped transmission line only produces linewidth modulation without resolvable mode splitting.

\newpage
\section*{Supplementary Note C: Magnon--Transmission Line Coupling}
The coupling strength between the YIG sphere and transmission line, $g_{mt}$, depends critically on the local electromagnetic field amplitude, which varies along the transmission line due to the standing wave pattern created by the cavity reflection and impedance discontinuities at the boundaries.

\subsection*{Spatial Coupling Profile Mapping}

We physically translated the YIG sphere along the microstrip line whilst monitoring the FMR response. This spatial scan directly maps the coupling strength $g_{mt}$ as a function of YIG position. 


The spacing between the nodes (positions where minimum RF field) and antinodes (positions where maximum field) can be calculated from the guided wavelength:
\[
\lambda_g = \frac{v_p}{f}, \qquad v_p = \frac{c}{\sqrt{\varepsilon_\text{eff}}},
\]
where $v_p$ is the phase velocity and $\varepsilon_\text{eff}$ is the effective permittivity.  
\begin{align*}
\text{Node $\leftrightarrow$ next node} &= \frac{\lambda_g}{2}, \\
\text{Node $\leftrightarrow$ adjacent antinode} &= \frac{\lambda_g}{4}.
\end{align*}


To determine key quantities such as the effective permittivity $\varepsilon_{\text{eff}}$ we can use the Hammerstad--Jensen~\cite{Hammerstad1980} -- a set of widely used analytic expressions for microstrip lines often implemented in many RF CAD tools.
For a microstrip line~\cite{pozar_microwave_2011},
\[
\varepsilon_\text{eff} = \frac{\varepsilon_r+1}{2} + \frac{\varepsilon_r-1}{2}\left(1 + \frac{12h}{w_\text{eff}}\right)^{-1/2}
\]
with effective width
\[
w_\text{eff} = 
\begin{cases}
w + \dfrac{t}{\pi}\!\left(1 + \ln \dfrac{4\pi w}{t}\right), & \text{if } w/h \leq 1, \\[1.2em]
w + \dfrac{t}{\pi}\!\left(1 + \ln \dfrac{2h}{t}\right), & \text{if } w/h > 1.
\end{cases}
\]



Below we compute $\lambda_g$ and spacings for our microstrip. The PCBs were fabricated by JLCPCB using standard FR4 substrate, which they specify as having $\varepsilon_r = 4.4$ at 1\,GHz. However, the relative permittivity of FR4 is known to decrease with frequency. 
At our operating frequency of 5.0\,GHz, we estimated $\varepsilon_r$ lies in the range 3.9--4.4, though the exact value is difficult to determine without dedicated dielectric characterization. 
For our calculations, we use $\varepsilon_r = 4.1$ as a reasonable estimate; we note that the final node-to-antinode spacing is relatively insensitive to this choice, varying by less than 6\% across the plausible range of permittivity values. 

The substrate thickness $h$ was determined by measuring the total PCB thickness with calipers, yielding 0.90--0.96\,mm from the bottom gold-plated ground plane to the top green solder mask surface. Subtracting the metal and coating layers---1\,$\mu$m bottom gold plating, 35\,$\mu$m bottom copper, and 20\,$\mu$m top solder mask---gives an effective dielectric height of $h \approx 0.874$\,mm. The trace width is $w = 1.943$\,mm and the total conductor thickness is $t = 36$\,$\mu$m (35\,$\mu$m copper plus 1\,$\mu$m gold plating on top). 
Using Hammerstad--Jensen:
\[
\varepsilon_{\text{eff}} \approx 3.17,\quad
\lambda_g \approx 33.69~\text{mm},\quad
\lambda_g/2 \approx 16.85~\text{mm},\quad
\lambda_g/4 \approx 8.42~\text{mm}.
\]
Nodes and antinodes should alternate every $\sim$8.42\,mm along the microstrip, which is in reasonable agreement with our experimental observations. For a YIG sphere positioned at distance $x_{\mathrm{YIG}}$ from an antinode, the spatial phase offset is therefore given by
\begin{equation}
\phi_0 (x) = \frac{360^\circ \cdot x_{\mathrm{YIG}}}{\lambda_g}.
\label{eq:supphi_0}
\end{equation}

\subsection*{Phase Shifter Control}

As established in Supplementary Note A, the doubled phase shift in our reflection geometry means that a 90$^\circ$ control setting produces a 180$^\circ$ standing wave shift, effectively moving antinodes to node positions. This is what provides precise control over YIG coupling without mechanical repositioning.


To verify this control mechanism, we performed ferromagnetic resonance (FMR) measurements on the YIG sphere whilst systematically varying the phase shifter setting (Figure~\ref{fig:TuningFMR}). These measurements directly probe the local electromagnetic field strength at the YIG position.

\begin{figure*}[htb]
\centering
\includegraphics[width=0.95\linewidth]{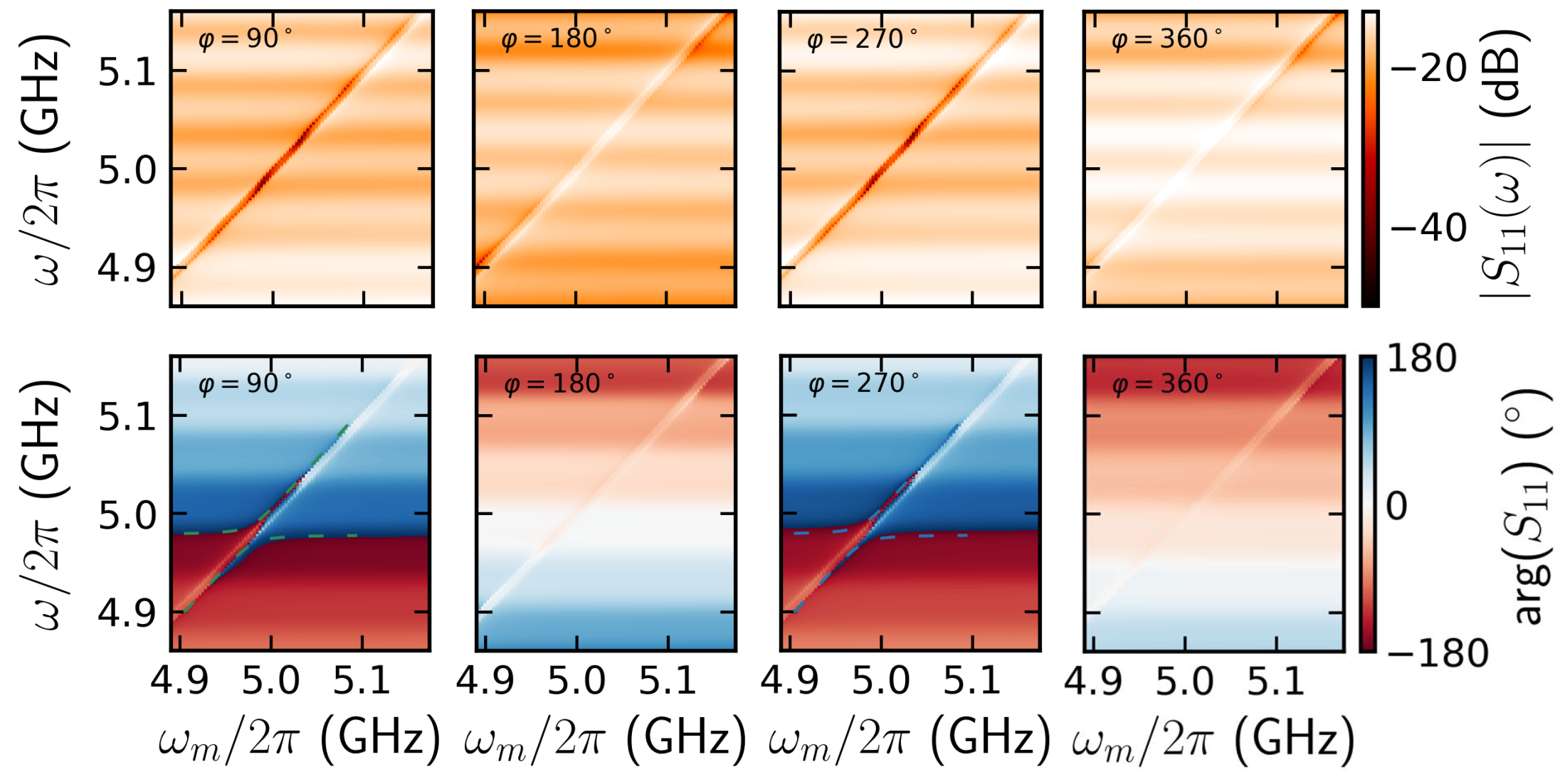}
\caption{FMR spectra showing YIG response as a function of phase shifter setting, demonstrating phase-dependent coupling modulation.}
\label{fig:TuningFMR}
\end{figure*} 

The FMR spectra confirm the theoretical prediction: the YIG coupling oscillates periodically with phase shifter setting, showing maximum coupling at odd multiples of 90$^\circ$ and complete decoupling (no observable FMR signal) at even multiples of 90$^\circ$ ($\varphi = 0^\circ, 180^\circ, 360^\circ, ...$). The $90^\circ$ periodicity in coupling corresponds to $\lambda/4$ shifts in the standing wave pattern, confirming the doubled-phase geometry. This phase-dependent modulation demonstrates continuous tunability of the magnon-photon coupling strength from maximum (YIG at antinode) to zero (YIG at node). From the phase data (more clear than amplitude data), we extract a maximum coupling strength of $g_{mt}^{\max}/2\pi \approx 10$~MHz for a YIG sphere of diameter 0.5~mm.

We can therefore express the magnon-transmission line coupling strength as:
\begin{equation}
g_{mt}(\varphi, x) = g_{mt}^{\mathrm{max}} \left|\sin(\varphi + \phi_0(x))\right|
\label{eq:supg_mt_phase}
\end{equation}
where $\varphi$ is the phase shifter setting and $\phi_0(x)$ is the spatial phase offset determined by the YIG position along the microstrip [Eq.~\ref{eq:supphi_0}]. 
Note that $g_{mt}^{\max}$ scales with $\sqrt{N_s}$, where $N_s$ is the total number of spins, allowing the coupling strength to be engineered by appropriate choice of sample size.



\newpage
\section*{Supplementary Note D: Time-Domain Transmission Measurements}

To measure the time-domain coupling dynamics, ringdown measurements were performed by applying a 16~ns square pulse at the fixed cavity frequency. The pulse was synthesized using an AMD RFSoC 4x2 development board; the pulse length was chosen empirically to optimize signal quality---16~ns provided the clearest ringdown signatures compared to shorter and longer pulses tested. 
The reflected signal was digitized using the same RFSoC board with a sampling rate of 614.4~MHz (1.63~ns per sample), providing sufficient temporal resolution to capture the coupling and decay behavior of interest ($\gamma_c \sim$ 1.68~MHz, $\gamma_m \sim$ 2~MHz, $g_{\mathrm{eff}} \sim$ 0.75~MHz). Each trace was averaged over 200 repetitions to improve signal-to-noise ratio.

The experimental data were processed by normalizing the measured ADC counts (proportional to voltage amplitude) to the maximum ADC value and converting to decibel scale:
\begin{equation}
S_{\text{exp}}(t) = 20 \log_{10}\!\left(\frac{V(t)}{V_{\text{max}}}\right)
                  = 20 \log_{10}\!\left(\frac{A(t)}{A_{\text{max}}}\right),
\end{equation}
where $A(t)$ represents the ADC value at time $t$

\subsection*{Theoretical Reflected Signal}
To obtain the theoretical time-domain response, we numerically integrate the equations of motion [Eqs.~\eqref{eq:cavity}–\eqref{eq:transmission}] using a fourth-order Runge–Kutta (RK4) with a time step of 1~ps. The input field $a_{\mathrm{in}}(t)$ is modeled as a 16~ns square pulse at the cavity frequency $\omega_c$, matching the experimental conditions. The reflected field is computed via the input–output relation [Eq.~\eqref{eq:suppinput_output}], and the reflected power normalized to the input power is:
\begin{equation}
S_{\text{theory}}(t)
 = 10 \log_{10}\!\left(\frac{|a_{\text{out}}(t)|^2}{|a_{\text{in,0}}|^2}\right)
 = 20 \log_{10}\!\left|\frac{a_{\text{out}}(t)}{a_{\text{in,0}}}\right|,
\label{eq:supS11_theory}
\end{equation}
where $a_{\text{in,0}}$ is the complex amplitude of the incident pulse. 
This definition is directly analogous to the experimentally measured $S_{11}$ parameter and automatically accounts for interference between the incident and re-radiated fields that produces dips or peaks in the reflected amplitude.

\subsubsection*{Mapping to Simulation Parameters}
The incident field amplitude $a_{\text{in,0}}$ is set by the microwave power applied at the input port. In the standard quantum optics convention, the input field amplitude relates to the photon flux through
\begin{equation}
|a_{\text{in,0}}|^2 = \frac{P_{\text{in}}}{\hbar \omega_c},
\end{equation}
where $|a_{\text{in,0}}|^2$ represents photon flux (photons per second). For our experiments at $f_c \sim 5$~GHz and $P_{\text{in}} \sim -20$ dBm
, this corresponds to $a_{\text{in,0}} \sim 2 \times 10^9$. 

\cleardoublepage
\section*{Supplementary Note E: Simplified Single Transmission-Line Mode Model}

Although the transmission line constitutes a continuum of propagating modes, the boundary conditions imposed by the cavity termination, finite line length, and any impedance mismatch create quasi-discrete standing wave resonances— we use these as our 'bad-cavity' auxiliary modes.

The coupling between the YIG sphere and the transmission line modes depends on the local field amplitude at the YIG position. When the YIG is located at a field antinode, coupling is maximized; when at a node, coupling is minimized. Changing the phase shifter setting alters the total electrical length of the transmission line path, which modifies the resonance conditions for standing wave modes, effectively shifting the positions of the standing wave nodes and antinodes along the transmission line. This creates a system where both the coupling strength to the YIG and the effective resonance frequencies of the transmission line can be controlled through the phase shifter.

\subsection*{Heisenberg equations (closed system)}
\begin{align}
\dot a &= -i\omega_c a - i g_{ct}\, t, \label{eq:supheisen_a}\\
\dot m &= -i\omega_m m - i g_{mt}\, t, \label{eq:supheisen_m}\\
\dot t &= -i\omega_{t} t - i g_{ct}\, a - i g_{mt}\, m. \label{eq:supheisen_t}
\end{align}

\subsection*{Open system: damping and input fields}
To model decay and measurements, the equations of motion become:
\begin{align}
\dot{a} &= -i\omega_c a - ig_{ct} t - \frac{\gamma_c}{2} a \label{eq:supcavity} \\
\dot{m} &= -i\omega_m m - ig_{mt} t - \frac{\gamma_m}{2} m \label{eq:supmagnon} \\
\dot{t} &= -i\omega_{t} t - ig_{ct} a - ig_{mt} m - \frac{\gamma_{t}}{2} t + \sqrt{\gamma_{\mathrm{ext}}}\, a_{\mathrm{in}}, \label{eq:suptransmission}
\end{align}
where $\gamma_c$, $\gamma_m$, $\gamma_{t}$ are the damping rates.
For the transmission-line mode, the total loss rate can be written as $\gamma_{t} = \gamma_{\mathrm{int}} + \gamma_{\mathrm{ext}}$, where $\gamma_{\mathrm{int}}$ accounts for intrinsic (internal) losses and $\gamma_{\mathrm{ext}}$ describes coupling to the external port through which the input field $a_{\mathrm{in}}$ and output field $a_{\mathrm{out}}$ are defined [see Fig.~\ref{fig:remoteCouplingSetUp}(b) of the main text].

\subsection*{Steady-state solution}

To find the steady-state response, we assume harmonic solutions at the drive frequency: $a = a_0 e^{-i\omega t}$, $m = m_0 e^{-i\omega t}$, $t = t_0 e^{-i\omega t}$, and $a_{\mathrm{in}} = a_{\mathrm{in}}^0 e^{-i\omega t}$. Substituting into the Heisenberg equations, we obtain three linear algebraic equations in frequency space:

\begin{align}
\Bigl(i(\omega_c - \omega) + \tfrac{\gamma_c}{2}\Bigr)a
  + i g_{ct}\, t &= 0, \label{eq:supafreq}\\
\Bigl(i(\omega_m - \omega) + \tfrac{\gamma_m}{2}\Bigr)m
  + i g_{mt}\, t &= 0, \label{eq:supmfreq}\\
\Bigl(i(\omega_t - \omega) + \tfrac{\gamma_t}{2}\Bigr)t
  + i g_{ct}\, a + i g_{mt}\, m &= \sqrt{\gamma_{\mathrm{ext}}}\, a_{\mathrm{in}}^0. \label{eq:suptfreq}
\end{align}
These equations can be written compactly in matrix form (after dividing by $i$):
\begin{equation}
\begin{pmatrix}
\Delta_c - i\frac{\gamma_c}{2} & 0 & g_{ct} \\
0 & \Delta_m - i\frac{\gamma_m}{2} & g_{mt} \\
g_{ct} & g_{mt} & \Delta_t - i\frac{\gamma_t}{2}
\end{pmatrix}
\begin{pmatrix}
a \\ m \\ t
\end{pmatrix}
=
\begin{pmatrix}
0 \\ 0 \\ -i\sqrt{\gamma_{\mathrm{ext}}}\, a_{\mathrm{in}}^0
\end{pmatrix}
\label{eq:supmatrix_system}
\end{equation}
where $\Delta_c = \omega_c - \omega$, $\Delta_m = \omega_m - \omega$, and $\Delta_t = \omega_t - \omega$.

\subsection*{\texorpdfstring{Reflection coefficient at the $t$ port ($S_{11}$)}{Reflection coefficient at the t port (S11)}}
To find the reflection coefficient, we solve the 3×3 matrix equation to obtain $t$, then substitute into the input-output relation. 
From the input--output relation: 
\begin{equation}
a_{\mathrm{out}} = a_{\mathrm{in}} - \sqrt{\gamma_{\mathrm{ext}}} t
\label{eq:suppinput_output}
\end{equation}
Therefore, the reflection coefficient is:
$$S_{11}(\omega) = \frac{a_{\mathrm{out}}}{a_{\mathrm{in}}} = 1 - \sqrt{\gamma_{\mathrm{ext}}} \frac{t}{a_{\mathrm{in}}}$$
Solving the matrix equation for $t$:
$$t = \frac{-i\sqrt{\gamma_{\mathrm{ext}}} a_{\mathrm{in}}^0}{\Delta_t - i\frac{\gamma_{t}}{2} - \frac{g_{ct}^2}{\Delta_c - i\frac{\gamma_c}{2}} - \frac{g_{mt}^2}{\Delta_m - i\frac{\gamma_m}{2}}}$$
Thus, the final scattering matrix element is:
\begin{equation}
S_{11}(\omega) = 1 - \frac{\gamma_{\mathrm{ext}}}{i\Delta_t + \frac{\gamma_{t}}{2} + \frac{g_{ct}^2}{i\Delta_c + \frac{\gamma_c}{2}} + \frac{g_{mt}^2}{i\Delta_m + \frac{\gamma_m}{2}}}
\label{eq:supS11}
\end{equation}

We can include the explicit phase dependence in this reduced single–mode model of Eq.~\eqref{eq:supS11} to obtain
\begin{equation}
S_{11}(\omega,\varphi,x)
= 1 - \frac{\gamma_{\mathrm{ext}}}{
i\Delta_t(\varphi) + \frac{\gamma_{t}}{2}
+ \dfrac{g_{ct}^2}{i\Delta_c + \frac{\gamma_c}{2}}
+ \dfrac{g_{mt}^2(\varphi,x)}{i\Delta_m + \frac{\gamma_m}{2}}
}.
\label{eq:supS11_phase}
\end{equation}
Here $\Delta_t(\varphi)=\omega_{t1}(\varphi)-\omega$ follows from Eq.~\eqref{eq:supomega_t1}, and the phase-dependent coupling is given by $g_{mt}(\varphi,x)$ from Eq.~\eqref{eq:supg_mt_phase}. This single–mode form is valid when only $\omega_{t1}(\varphi)$ participates and $\omega_{t2}(\varphi)$ remains far detuned, which holds for $\varphi\simeq 0^\circ$–$180^\circ$.


\subsection*{\texorpdfstring{Eliminating the bus mode $t$}{Eliminating the bus mode t}}

It is sometimes convenient to eliminate $t$ algebraically to investigate the effective interaction between the cavity and the magnon. Solving Eq.~\eqref{eq:suptfreq} for $t$ and substituting back into Eqs.~\eqref{eq:supafreq}--\eqref{eq:supmfreq} yields a $2\times 2$ system for $(a,m)$ with an effective coupling.

From Eq.~\eqref{eq:suptfreq}:
\begin{equation}
t = \frac{-i\sqrt{\gamma_t} a_{\mathrm{in}}^0 - g_{ct} a - g_{mt} m}{\Delta_t - i\frac{\gamma_t}{2}}
\label{eq:supt_eliminated}
\end{equation}
Substituting this into Eqs.~\eqref{eq:supafreq} and \eqref{eq:supmfreq}:
\begin{align}
\left[-\Delta_c + i\frac{\gamma_c}{2} + \frac{g_{ct}^2}{\Delta_t - i\frac{\gamma_t}{2}}\right] a
+ \frac{g_{ct} g_{mt}}{\Delta_t - i\frac{\gamma_t}{2}} m
&= \frac{g_{ct}\sqrt{\gamma_t}}{i\Delta_t + \frac{\gamma_t}{2}}\, a_{\mathrm{in}}^0, \label{eq:supa_effective}\\
\frac{g_{ct} g_{mt}}{\Delta_t - i\frac{\gamma_t}{2}} a
+ \left[-\Delta_m + i\frac{\gamma_m}{2} + \frac{g_{mt}^2}{\Delta_t - i\frac{\gamma_t}{2}}\right] m
&= \frac{g_{mt}\sqrt{\gamma_t}}{i\Delta_t + \frac{\gamma_t}{2}}\, a_{\mathrm{in}}^0. \label{eq:supm_effective}
\end{align}

Eliminating the transmission-line mode from Eqs.~\eqref{eq:supafreq}–\eqref{eq:suptfreq} yields a reduced two-mode model for the cavity and magnon amplitudes, in which the transmission line appears implicitly through frequency-dependent self-energy and coupling terms.
The equations for $(a,m)$ can be written compactly to reveal the effective magnon--cavity coupling:
\begin{equation}
\label{eq:supam-matrix}
\begin{pmatrix}
z_c(\omega) & g_{\mathrm{eff}}(\omega) \\[6pt]
g_{\mathrm{eff}}(\omega) & z_m(\omega)
\end{pmatrix}
\begin{pmatrix} a \\ m \end{pmatrix}
=
\frac{\sqrt{\gamma_t}}{i\Delta_t + \frac{\gamma_t}{2}}
\begin{pmatrix} g_{ct} \\ g_{mt} \end{pmatrix}
a_{\mathrm{in}}^{\,0}.
\end{equation}
with
\begin{equation}
z_n(\omega)
= -\Delta_n + i\frac{\gamma_n}{2}
+ \frac{g_{nt}^2}{\Delta_t - i\frac{\gamma_t}{2}},
\quad n\in\{c,m\}.
\end{equation}

In the remote coupling setup, the cavity and magnon modes do not interact directly but couple through the transmission line as an intermediary. This creates an \textit{effective coupling} between the cavity and magnon given by:
\begin{equation}
g_{\mathrm{eff}}(\omega)
= \frac{g_{ct} g_{mt}}{\Delta_t - i\frac{\gamma_t}{2}}.
\label{eq:supg_eff}
\end{equation}


The effective coupling between the cavity and magnon given in Eq.~\ref{eq:supg_eff} is inherently complex and frequency-dependent. This contrasts with direct coherent coupling (which would be purely real) and reveals the dissipative nature of the interaction. The coupling can be decomposed into coherent and dissipative components:
\begin{equation}
g_{\mathrm{eff}}(\omega) = g_{\mathrm{coh}}(\omega) + i\Gamma_{\mathrm{diss}}(\omega),
\end{equation}
where
\begin{align}
g_{\mathrm{coh}}(\omega) &= \frac{g_{ct} g_{mt} \Delta_t}{\Delta_t^2 + (\gamma_t/2)^2}, \label{eq:supcoherent_part}\\
\Gamma_{\mathrm{diss}}(\omega) &= \frac{g_{ct} g_{mt} (\gamma_t/2)}{\Delta_t^2 + (\gamma_t/2)^2}. \label{eq:supdissipative_part}
\end{align}

\subsubsection*{Determining the eigenmodes}

The eigenmodes of the coupled cavity--magnon system correspond to the natural oscillation frequencies and can be determined from the homogeneous version of Eq.~\eqref{eq:supam-matrix}. These modes describe how the system oscillates freely in the absence of external driving.

For free oscillations, we set the driving term to zero ($a_{\mathrm{in}}^0 = 0$) and seek non-trivial solutions to:
\begin{equation}
\begin{pmatrix}
z_c(\omega) & g_{\mathrm{eff}}(\omega) \\[12pt]
g_{\mathrm{eff}}(\omega) & z_m(\omega)
\end{pmatrix}
\begin{pmatrix} a \\ m \end{pmatrix} = 0,
\label{eq:supeigenvalue_problem}
\end{equation}
Non-trivial solutions exist when the determinant of the matrix vanishes:
\begin{equation}
\det[\mathbf{Z}(\omega)] = z_c(\omega)\, z_m(\omega) - g_{\mathrm{eff}}^2(\omega) = 0.
\label{eq:supeigenmode_condition}
\end{equation}



In the strongly damped limit ($\gamma_t/2 \gg |\Delta_t|, g_{ct}, g_{mt}$) the transmission-line mode can be adiabatically eliminated, making the frequency dependence of $g_{\mathrm{eff}}(\omega)$ and the self-energy terms negligible. In this regime the effective parameters become approximately constant, and the eigenvalue problem can be recast in the standard form 
$\omega \mathbf{v} = \mathbf{H}_{\mathrm{eff}}\mathbf{v}$, where $\omega$ appears explicitly as the eigenvalue. The effective Hamiltonian is:
\begin{equation}
\mathbf{H}_{\mathrm{eff}} = 
\begin{pmatrix} 
\omega_c - i\gamma_c'/2 & i\Gamma \\[4pt]
i\Gamma & \omega_m - i\gamma_m'/2
\end{pmatrix},
\end{equation}

where the renormalized damping rates include contributions from the transmission line:
\begin{align}
\gamma_c' &= \gamma_c + \frac{2g_{ct}^2}{\gamma_t}, \\
\gamma_m' &= \gamma_m + \frac{2g_{mt}^2}{\gamma_t},
\end{align}
and the dissipative coupling strength is:
\begin{equation}
\Gamma = \frac{2g_{ct}g_{mt}}{\gamma_t}.
\end{equation}

The eigenfrequencies can also be found analytically by solving the characteristic equation $\det(\mathbf{H}_{\mathrm{eff}} - \omega\mathbf{I}) = 0$:
\begin{equation}
\omega_{\pm} = \bar{\omega} - i\bar{\gamma} \pm \sqrt{\left(\frac{\Delta}{2} - i\frac{\delta\gamma}{4}\right)^2 - \Gamma^2},
\end{equation}
where $\bar{\omega} = (\omega_c + \omega_m)/2$ is the average frequency, $\bar{\gamma} = (\gamma_c' + \gamma_m')/4$ is the average damping, $\Delta = \omega_c - \omega_m$ is the cavity--magnon detuning, and $\delta\gamma = \gamma_c' - \gamma_m'$ is the damping difference.


We note that following Ref. \cite{Zhao2020}, this Hamiltonian can be cast into anti-$\mathcal{PT}$-symmetric form. Transforming into the frame rotating at $\bar{\omega}$ the effective Hamiltonian becomes 

\begin{equation}
H' = H_{\mathrm{eff}} - \bar{\omega} I
=
\begin{pmatrix}
\frac{\Delta}{2} - i\gamma_c'/2 & i\Gamma \\[6pt]
i\Gamma & -\frac{\Delta}{2} - i\gamma_m'/2
\end{pmatrix}.
\label{eq:Hprime_basic}
\end{equation}
This effective Hamiltonian is anti-$\mathcal{PT}$-symmetric when $\gamma_c' = \gamma_m'$.

\cleardoublepage
\section*{Supplementary Note F: Full Model With Two Transmission-Line Modes}

The microstrip transmission line supports multiple standing-wave resonances that can simultaneously couple to both the cavity and magnon modes (see Supplementary Note~B). Experimentally, two transmission-line modes cross the cavity resonance as the phase shifter setting is varied. Their phase-dependent frequencies follow Eqs.~\eqref{eq:supomega_t1}–\eqref{eq:supomega_t2}. Because both transmission-line modes share the same input/output port and couple in the same manner to the cavity and magnon, the full dynamical system consists of four coupled modes.

\subsection*{Heisenberg Equations of Motion for the four-mode model}

The Heisenberg–Langevin equations for this four-mode system are:
\begin{align}
\dot{a} &= 
\left(-i\omega_c - \frac{\gamma_c}{2}\right)a
- i g_{ct}(t_1 + t_2), 
\label{eq:supcavity_twomode} \\
\dot{m} &= 
\left(-i\omega_m - \frac{\gamma_m}{2}\right)m
- i g_{mt}(t_1 + t_2),
\label{eq:supmagnon_twomode} \\
\dot{t}_1 &= 
\left(-i\omega_{t1} - \frac{\gamma_t}{2}\right)t_1
- i g_{ct} a 
- i g_{mt} m
+ \sqrt{\gamma_{\mathrm{ext}}}\, a_{\mathrm{in}},
\label{eq:supt1} \\
\dot{t}_2 &= 
\left(-i\omega_{t2} - \frac{\gamma_t}{2}\right)t_2
- i g_{ct} a 
- i g_{mt} m
+ \sqrt{\gamma_{\mathrm{ext}}}\, a_{\mathrm{in}}.
\label{eq:supt2}
\end{align}
Since both transmission-line modes are driven by the same input field, the input–output relation becomes:
\begin{equation}
a_{\mathrm{out}} = a_{\mathrm{in}} - \sqrt{\gamma_{\mathrm{ext}}}\,(t_1 + t_2).
\label{eq:supIO_twomode}
\end{equation}
This full four-mode model captures all features of the system dynamics. The simplified single–transmission-line-mode model used in the main text is therefore a reduced version appropriate for illustrating the essential physics.

\subsection*{Steady-state solution and reflection coefficient for the four-mode model}

To obtain the steady-state response of the full four-mode system, we again assume harmonic time dependence at the drive frequency: 
\[
a = a_0 e^{-i\omega t},\quad
m = m_0 e^{-i\omega t},\quad
t_j = t_{j0} e^{-i\omega t}\ (j=1,2),\quad
a_{\mathrm{in}} = a_{\mathrm{in}}^0 e^{-i\omega t}.
\]


Substituting into Eqs.~\eqref{eq:supcavity_twomode}–\eqref{eq:supt2} yields the algebraic equations
\begin{align}
\bigl(i\Delta_c + \tfrac{\gamma_c}{2}\bigr)a 
+ ig_{ct}(t_1 + t_2) &= 0, \\
\bigl(i\Delta_m + \tfrac{\gamma_m}{2}\bigr)m 
+ ig_{mt}(t_1 + t_2) &= 0, \\
\bigl(i\Delta_{t1} + \tfrac{\gamma_t}{2}\bigr)t_1 
+ ig_{ct} a + ig_{mt} m &= \sqrt{\gamma_{\mathrm{ext}}}\, a_{\mathrm{in}}^0, \\
\bigl(i\Delta_{t2} + \tfrac{\gamma_t}{2}\bigr)t_2 
+ ig_{ct} a + ig_{mt} m &= \sqrt{\gamma_{\mathrm{ext}}}\, a_{\mathrm{in}}^0,
\end{align}
where $\Delta_c = \omega_c - \omega$, $\Delta_m = \omega_m - \omega$, and $\Delta_{tj} = \omega_{tj} - \omega$ for $j=1,2$.

The first two equations depend only on the sum $t_1 + t_2$. Solving them gives
\begin{align}
a &= -\frac{ig_{ct}}{i\Delta_c + \frac{\gamma_c}{2}}\,(t_1 + t_2), \\
m &= -\frac{ig_{mt}}{i\Delta_m + \frac{\gamma_m}{2}}\,(t_1 + t_2).
\end{align}

Substituting these into the equations for $t_1$ and $t_2$ and solving the resulting $2\times 2$ system yields the sum of the transmission-line amplitudes
\begin{equation}
t_1 + t_2 
= \sqrt{\gamma_{\mathrm{ext}}}\, a_{\mathrm{in}}^0 \,
\frac{i(\Delta_{t1} + \Delta_{t2}) + \gamma_t}{
\bigl(i\Delta_{t1} + \tfrac{\gamma_t}{2}\bigr)\bigl(i\Delta_{t2} + \tfrac{\gamma_t}{2}\bigr)
+ \bigl[i(\Delta_{t1} + \Delta_{t2}) + \gamma_t\bigr]
\left(
\frac{g_{ct}^2}{i\Delta_c + \frac{\gamma_c}{2}}
+ \frac{g_{mt}^2}{i\Delta_m + \frac{\gamma_m}{2}}
\right)
}.
\label{eq:sup_tsum_twomode}
\end{equation}

Since both transmission-line modes couple to the same port, the input--output relation generalises to
\begin{equation}
a_{\mathrm{out}} = a_{\mathrm{in}} - \sqrt{\gamma_{\mathrm{ext}}}\,(t_1 + t_2),
\end{equation}
so that the reflection coefficient at the port is
\begin{equation}
S_{11}(\omega) 
= \frac{a_{\mathrm{out}}}{a_{\mathrm{in}}}
= 1 - \sqrt{\gamma_{\mathrm{ext}}}\,\frac{t_1 + t_2}{a_{\mathrm{in}}^0}.
\end{equation}


Using Eq.~\eqref{eq:sup_tsum_twomode}, we obtain the final expression
\begin{equation}
S_{11}(\omega)
= 1 - \gamma_{\mathrm{ext}} \,
\frac{i(\Delta_{t1} + \Delta_{t2}) + \gamma_t}{
\bigl(i\Delta_{t1} + \tfrac{\gamma_t}{2}\bigr)\bigl(i\Delta_{t2} + \tfrac{\gamma_t}{2}\bigr)
+ \bigl[i(\Delta_{t1} + \Delta_{t2}) + \gamma_t\bigr]
\left(
\frac{g_{ct}^2}{i\Delta_c + \frac{\gamma_c}{2}}
+ \frac{g_{mt}^2}{i\Delta_m + \frac{\gamma_m}{2}}
\right)
}.
\label{eq:supS11_twomode}
\end{equation}
This expression reduces to the single–transmission-line-mode result [Eq.~\eqref{eq:supS11}] in the limit where only a single transmission-line mode is relevant, i.e.\ when the second mode is far detuned ($\Delta_{t2} \to \infty$).

To capture the full phase-dependent behaviour, we write Eq.~\eqref{eq:supS11_twomode} explicitly as
\begin{small}
\begin{equation}
S_{11}(\omega,\varphi,x)
= 1 - \gamma_{\mathrm{ext}} \,
\frac{i[\Delta_{t1}(\varphi) + \Delta_{t2}(\varphi)] + \gamma_t}{
\bigl(i\Delta_{t1}(\varphi) + \tfrac{\gamma_t}{2}\bigr)\bigl(i\Delta_{t2}(\varphi) + \tfrac{\gamma_t}{2}\bigr)
+ \bigl[i[\Delta_{t1}(\varphi) + \Delta_{t2}(\varphi)] + \gamma_t\bigr]
\left(
\frac{g_{ct}^2}{i\Delta_c + \frac{\gamma_c}{2}}
+ \frac{g_{mt}^2(\varphi,x)}{i\Delta_m + \frac{\gamma_m}{2}}
\right)
}.
\label{eq:supS11_twomode_phase}
\end{equation}
\end{small}
Here $\Delta_{tj}(\varphi) = \omega_{tj}(\varphi) - \omega$ for $j = 1,2$, with the phase-dependent transmission-line frequencies given by Eqs.~\eqref{eq:supomega_t1} and \eqref{eq:supomega_t2}, and the phase-dependent coupling $g_{mt}(\varphi,x)$ given by Eq.~\eqref{eq:supg_mt_phase}. Incorporating both phase-dependent transmission-line modes captures additional structure in the reflection spectrum, enabling more realistic modelling of the experimental system across the full range of phase shifter settings.

\endgroup

\end{document}